\documentclass[11pt,letterpaper]{article}
\usepackage[margin=1.05in]{geometry}
\usepackage{amsmath}
\usepackage{setspace}
\usepackage{authblk}
\usepackage{microtype}
\usepackage{hyperref}
\begin{document}
\title{Resurrecting the Strong KSS Conjecture}
\author{Scott Lawrence\footnote{scott.lawrence-1@colorado.edu}}
\affil{Department of Physics, University of Colorado, Boulder, CO 80309, USA}
\maketitle
\vspace{-1.3em}
\begin{abstract}
Many counterexamples to the proposed KSS bound $\frac\eta s \ge \frac 1 {4\pi}$ depend on constructing systems with large numbers of species. As a result, the entropy density grows large and $\frac \eta s$ can be made arbitrarily small. However, these constructions do not affect the dimensionless shear viscosity $\frac {\eta T}{\epsilon + P}$, which agrees with the traditional $\frac \eta s$ only for vanishing chemical potential. This raises the possibility that a KSS-like bound holds for all systems, not just UV-complete quantum field theories, contrary to the previous understanding.
\end{abstract}

In 2005, Kovtun, Son, and Starinets proposed, on the basis of both holographic and physical examples, a universal lower bound for the ratio of the shear viscosity to entropy density~\cite{Kovtun:2004de}: $\frac{\eta}{s} \ge \frac 1 {4\pi}$. All known physical systems obey this bound, with the quark-gluon plasma coming closest~\cite{Schafer:2009dj}. Since then, apparent counterexamples have been constructed via the AdS/CFT correspondence~\cite{Cremonini:2011iq}. However, in no such holographic construction has $\frac \eta s$ been made arbitrarily small without compromising the consistency of the theory. This suggests at least that a generalized bound $\frac \eta s \ge C_{\mathrm{KSS}}$, for some universal constant $C_{\mathrm{KSS}}$, may still hold.

What is the family of systems that must obey such a bound? Originally it was hoped that the KSS bound might hold for all systems described by hydrodynamics at large time and distance scales, or at least all such quantum mechanical systems (the \emph{strong KSS conjecture}). Soon thereafter, however, a series of counterexamples were constructed~\cite{Cohen:2007qr,Cherman:2007fj} that appeared to dramatically narrow the possible scope of this bound or any similar bound $\frac{\eta}{s} \ge C_{\mathrm{KSS}}$. What these examples all have in common is the introduction of a large number of species $N_s$. At least in the case of a dilute gas, this can be done without changing the dynamics of the system, including $\eta$. The associated entropy of mixing contributes a term $\log N_s$ to the entropy density. Thus, by taking a sufficiently large value of $N_s$, the ratio $\frac{\eta}{s}$ can be made arbitrarily small, violating any generalized KSS bound. After the consideration of these examples, \cite{Cherman:2007fj} proposed that a generalized KSS bound could be expected at most to hold for UV-complete quantum field theories, and only at vanishing chemical potential. Notably, this weakened KSS conjecture excludes many of the systems that were the original motivation.

For concreteness, I will sketch one such construction in some detail. Consider a nearly free (dilute) gas, with mean free path $\lambda$, mean velocity $\bar v$, density $\rho$, and $N_s$ species of equal mass. As long as $\lambda$ is long relative to all other scales, we can reliably calculate the shear viscosity from $\eta \approx C \lambda \rho \bar v$. (Here $C$ is a dimensionless, order-one constant.) Although the shear viscosity does not depend on $N_s$, the entropy density does. In the same limit, we have
\begin{equation}
s \approx n\left[\log N_s + \frac 5 2 + \log \frac{\bar v^{3/2}}{n}\right]
\text,
\end{equation}
where $n$ is the total number density.
The first term is the entropy of mixing. Without changing any other physical properties of the system, we can take $N_s$ to be arbitrarily large, making $\frac \eta s$ as close to $0$ as desired.

Here we see a small paradox. Although $\frac \eta s$ is made arbitrarily small as we increase $N_s$, nothing about the dynamics of the system has been changed. In particular, in (relativistic) viscous hydrodynamics, the long-time behavior of a shear wave is an exponential decay~\cite{Romatschke:2017ejr}:
\begin{equation}
\big\langle T^{01}(k,t) T^{01}(k,0)\big \rangle
\sim
e^{- \frac{\eta}{\epsilon + P}k^2 t}
\text.
\end{equation}
The rate of decay of these shear waves is not modified by the above construction, although $\frac \eta s$ becomes arbitrarily small. (Above, $T^{\mu\nu}$ is the energy-momentum tensor, meaning that $T^{01}$ represents momentum density in the $x$-direction. The wavenumber $k$ is lies in the $y$-$z$ plane.)

To resolve this paradox, consider the physical meaning of $\frac \eta s$. Ordinarily, the exponential decay motivates the study of the ratio $\frac{\eta}{s}$ as follows. The Gibbs free energy density is given by $g = \epsilon + P - sT$.
In a system at vanishing chemical potential $\mu = 0$, the Gibbs free energy also vanishes, so that $sT = \epsilon + P$. The rate of decay can then be written $\frac \eta s \frac {k^2}{T}$ -- note the appearance of $\frac \eta s$. However, the construction above is at $\mu > 0$, breaking the relationship between $\frac \eta s$ and the decay of shear waves. This is how $\frac \eta s$ can become arbitrarily small while shear waves still decay quickly.

A different dimensionless ratio, $\hat\eta \equiv \frac{\eta T}{\epsilon + P}$, is relevant at all values of $\mu$ -- I will refer to this as the \emph{dimensionless (shear) viscosity}. The decay rate of a shear wave is given by $\hat \eta \frac{k^2}{T}$. The physical relevance of the dimensionless shear viscosity is not limited to shear waves: it also appears in the decay rate of sound waves, accompanied by a similarly constructed dimensionless bulk viscosity. A modified KSS bound, $\hat \eta \ge C_{\mathrm{KSS}}$, is not broken by any construction like the one above, as the entropy of mixing does not appear in the dimensionless shear viscosity.

In~\cite{Cherman:2007fj} it was suggested that the distinguishing feature of the large-$N_s$ construction was that it could not easily be performed in a quantum field theory, and therefore that a generalized KSS bound could not be expected to hold without UV-completeness. In light of the above discussion, we can now take the view that instead, the distinguishing feature of that theory was $\mu > 0$. When working with the dimensionless viscosity $\hat \eta$, which remains relevant for the dynamics of the theory even at nonvanishing chemical potential, this feature does not matter. This suggests that a generalized KSS bound $\hat \eta \ge C_{\mathrm{KSS}}$ may hold for all systems described at large distances and times by hydrodynamics.

What could be behind such a universal bound? A prototype of an argument in this direction is given in~\cite{Kovtun:2011np}. The authors of that paper show that, when shear and sound waves fail to decay quickly, they become an effective mechanism for the transport of momentum, and have their own contribution to the dimensionless viscosity. As a result, in a system consistently described by second-order relativistic hydrodynamics, the decay of sound and shear waves cannot be too slow; i.e., the dimensionless shear viscosity is bounded from below.

A variety of the proposed holographic counterexamples to the original KSS bound are still applicable to the modification proposed here; see~\cite{Cremonini:2011iq} for a review. To my knowledge, no counterexample has been constructed that allows for arbitrarily small $\hat \eta$ while preserving the consistency of the theory.

\emph{Acknowledgements.} I am grateful to Tom Cohen, Paul Romatschke, and Yukari Yamauchi for helping me to understand emergent hydrodynamics. Henry Lamm, Brian McPeak, Paul Romatschke, and Yukari Yamauchi gave valuable feedback on an earlier draft of this manuscript. This work was supported by the U.S. Department of Energy under Contract No. DE-SC0017905.

\bibliographystyle{unsrt}
\bibliography{kss}
\end{document}